\documentclass[conference]{IEEEtran}
\IEEEoverridecommandlockouts

\usepackage{cite}
\usepackage{amsmath,amssymb,amsfonts}
\usepackage{algorithmic}
\usepackage{graphicx}
\usepackage{subfigure}
\usepackage{textcomp}
\usepackage{url}
\usepackage{caption}
\usepackage{tabularx}
\usepackage{makecell}
\usepackage{booktabs}
\usepackage{diagbox}
\usepackage{algorithm}
\usepackage{xcolor}
\def\BibTeX{{\rm B\kern-.05em{\sc i\kern-.025em b}\kern-.08em
    T\kern-.1667em\lower.7ex\hbox{E}\kern-.125emX}}
\begin{document}

\title{GalSAS-SDR-SIM: An End-to-End Simulation Platform for Galileo Signal Authentication Service}
\author{Haiyang Wang, 
	Yuanyu Zhang,
	Wensen Du, 
	Ji He, 
	Pinchang Zhang,
	Ning Xi
	\thanks{This work was supported in part by the National Natural Science Foundation of China (No. 62220106004, 92467201), in part by the Fundamental Research Funds for the Central Universities (Grant No. QTZX25080, ZDRC2202), in part by Shaanxi Elite Talent Introduction Program (Youth Project), in part by the Young Talent Fund of Association for Science and Technology in Shaanxi, China. 
    (\emph{Corresponding author}: Yuanyu Zhang)}
	\thanks{H. Wang, Y. Zhang, W. Du, J. He, and N. Xi are with the School of Computer Science \& Technology, Xidian University, Xi'an, Shaanxi 710071, China. P. Zhang is with the College of Cybersecurity, Tarim University, Xinjiang, China. Email: \{hywang\_xdu, 25031212258\}@stu.xidian.edu.cn, \{yyuzhang, jihe, nxi\}@xidian.edu.cn, 120260003@taru.edu.cn.}
}
\maketitle

\begin{abstract}
Galileo is developing a Signal Authentication Service (SAS) that integrates Open Service Navigation Message Authentication (OSNMA) on the E1 band with Spreading Code Authentication (SCA) on the E6 band to strengthen its resilience to spoofing attacks. 
As Galileo SAS is still under development, access to realistic and controllable SAS signals remains limited, hindering both the early development of compatible receivers and reproducible research on signal authentication.
To bridge this gap, this paper presents GalSAS-SDR-SIM, an open-source software-defined radio (SDR) simulation platform that emulates the SAS workflow by coupling E6 code encryption with the OSNMA key-disclosure process. 
The platform allows flexible SAS configuration of code encryption parameters to accommodate receivers with different computational capabilities. 
It also supports the concurrent generation of Galileo E1, E5b, and E6 signals for user-defined locations and times, and OSNMA cross-satellite configurations. 
Experimental results demonstrate simultaneous verification of navigation messages and spreading codes.
We further evaluate SAS authentication performance and computational resource costs under different SAS configurations.
Implemented according to publicly available official specifications, GalSAS-SDR-SIM provides a practical tool for accelerating SAS-capable receiver development and supporting the research community in evaluating and improving Galileo signal-authentication techniques.
\end{abstract}


\section{Introduction}
Global Navigation Satellite Systems (GNSS) have become a crucial foundation for intelligent transportation, unmanned systems, and logistics. Because traditional civilian GNSS signal formats and modulation methods are publicly available and lack encryption mechanisms, attackers can forge signals, deceiving GNSS receivers into calculating position and time solutions \cite{spoof6-usenix-cars,uavtakover-usenix54}. Furthermore, successful spoofing attacks can propagate to upper-layer network services that rely on trusted time and location information, including vehicular communications, power grid systems, and financial systems \cite{noah2025gnss-survey,tibaldo2025gnss}.

Galileo was the first navigation satellite system to introduce Open Service Navigation Message Authentication (OSNMA) for civilian users \cite{damy2024increasing,llorca2023study}. 
Its OSNMA authenticates Galileo E1-B navigation data using the delayed key disclosure protocol and message authentication codes (MACs) \cite{simon2025galileo,o2025galileo}. 
OSNMA substantially raises the bar for message-level spoofing because the attacker cannot forge valid MACs before the corresponding keys are disclosed \cite{hammarberg2024experimental,rusu2025experimental}. 
However, OSNMA cannot authenticate spreading codes or ranging measurements. 
Attackers can still use signal-level attacks to manipulate Galileo code phase while retaining valid navigation data to achieve spoofing \cite{replay12,lazyreplay2023,wang2023novel52}. 
The Galileo Signal Authentication Service (SAS) is being developed to address this gap by combining message authentication on E1 with spreading code authentication (SCA) on E6 \cite{fernandez2023semiassisted,fernandez2024galileo}. 
In the SAS, selected fragments of the encrypted E6-C code signal are re-encrypted with keys in the OSNMA and published as Re-Encrypted Code Sequences (RECS). 
A receiver downloads the RECS and related metadata in advance, records E6-C snapshots when the corresponding encrypted code fragments are transmitted, and later decrypts the RECS after the OSNMA key is disclosed. 
By correlating the recovered encrypted code sequence with the stored E6-C samples, the receiver obtains an a posteriori check on the signal component that produced the ranging measurement. 
In this way, OSNMA and SCA can provide the authenticated navigation data and authenticated ranging evidence from which a receiver may construct an authenticated position, velocity, and time (PVT) solution.
SAS services are expected to significantly enhance the security of Galileo navigation signals \cite{ramirez2025looking}.

\textbf{Motivation:} Despite its promising future, SAS research currently faces a practical bottleneck: a lack of publicly available end-to-end testing environments.
Operational SAS signals are not yet widely available to researchers and receiver developers, and public signal-in-space campaigns are limited in time, geometry, and configuration.
Researchers cannot repeatedly adjust receiver locations and times, satellite visibility, E1/E6 channel quality, OSNMA configuration, SAS configuration, such as SCA sequence lengths, authentication frequency and windows.
Therefore, receiver manufacturers cannot develop SAS-enabled receivers in advance, and thoroughly evaluate their performance and compare different receiver strategies, until SAS services are officially operational. 
This also hinders the research community from conducting performance evaluations, security analyses and optimizations for SAS.

To address the gap, this paper presents GalSAS-SDR-SIM, an open-source software-defined radio (SDR) simulation platform for research on Galileo SAS \footnote{https://github.com/Haiyang-Wong/GalSAS-SDR-SIM}. 
The platform implements the public SAS workflow by coupling E6 code encryption with the OSNMA key-disclosure process.
The platform allows flexible SAS configuration of code encryption parameters to accommodate receivers with different computational capabilities. 
In addition, it also supports the concurrent generation of Galileo E1, E5b, and E6 signals for user-defined locations and times, OSNMA cross-satellite authentication, and signal transmission via SDR hardware.
The platform supports receiver-side verification that combines OSNMA authentication with E6-C code authentication.
This platform does not claim to be an official implementation of the final operational SAS service, as its public parameters may still evolve. 
Instead, it provides a reproducible and extensible experimental tool that follows publicly available SAS principles and supports receiver prototyping, security analysis, and protocol exploration.
The main contributions are as follows.
\begin{itemize}
    \item We design an end-to-end SAS research platform that jointly simulates Galileo E1, E5b, and E6 signals, integrates OSNMA message authentication with E6-C spreading code authentication, and supports both offline baseband generation and SDR-based over-air transmission.
    \item We implement configurable OSNMA and SAS simulations. The platform allows configuration of RECS generation, scheduling and authentication cycle, OSNMA cross-authentication, correlation-based SCA verification, and scenario control over time, location, and satellite visibility.
    \item We build a receiver-side evaluation workflow that verifies OSNMA-authenticated navigation data and authenticated E6-C code measurements.  
    We conducted extensive experiments under different SAS configurations to analyze and evaluate aspects including authentication correctness, authentication latency, sensitivity to SAS parameters, and computational resource overhead.
\end{itemize}

The remainder of the paper is organized as follows. Section~\ref{sec:background} summarizes the SAS mechanism. Section~\ref{sec:design} presents the design and implementation of GalSAS-SDR-SIM. Section~\ref{sec:evaluation} describes the experimental methodology and evaluation results. Section~\ref{sec:conclusion} concludes the paper.

\section{SAS Background and Mechanism}\label{sec:background}
\subsection{GNSS Authentication Goals}
GNSS navigation satellite broadcast signals contain navigation messages and pseudo-random noise (PRN) codes. 
The navigation message includes satellite clock and orbit information, while the spreading code phase can be used to estimate pseudorange measurements. 
GNSS receivers receive this signal and parse the navigation message and code phase to calculate the PVT solution. 
Because traditional GNSS signals do not encrypt the navigation message or the PRN code, attackers can simultaneously forge the navigation message and adjust the code phase to modify the pseudorange information, spoofing the receiver to arbitrary locations and times.

The Galileo navigation system first uses OSNMA to verify navigation messages. 
OSNMA uses delayed TESLA keys to generate MACs for the navigation message. 
The receiver collects the broadcast navigation data and MAC, waits for the delayed disclosure of the corresponding key to calculate the local MAC, and then checks if the received MAC matches the calculated MAC. 
If they match, the received navigation message is authentic. 
However, OSNMA does not prevent message-level replay attacks, which first forge the code phase and Doppler frequency but retain the original navigation message, then remodulate and retransmit the signal. 
Because the signal contains authentic OSNMA data, it can still pass the receiver's OSNMA authentication and spoof the receiver's position and time \cite{lazyreplay2023,distributereplay40}. 
Many recent attacks against OSNMA exploit this idea.

The Galileo system further introduced SAS service, designed to supplement OSNMA by authenticating the code used for ranging. 
SAS uses the SCA mechanism to prevent attackers from directly forging pseudorange information by encrypting the spreading code.
If both the navigation message and pseudorange information received by the receiver are authenticated, the receiver can use them as authenticated inputs to its PVT engine.

\subsection{OSNMA}\label{sec:OSNMA}
OSNMA authenticates Galileo E1-B I/NAV data through message authentication tags and delayed TESLA key disclosure. 
Each E1-B page contains a 40-bit OSNMA field, which is divided into HKROOT and MACK parts. HKROOT carries the signed root-key and configuration information, while MACK carries authentication tags and disclosed TESLA keys \cite{OSNMAsisicd20,OSNMAreceiver21,perrig2003tesla}.

In OSNMA, the Galileo satellites disclose TESLA keys only after the authenticated E1 navigation data have already been broadcast. A receiver first verifies the key chain from the signed root key, then uses the disclosed key to check previously received tags \cite{gotzelmann2023OSNMA58}.
OSNMA uses the Authentication Data and Key Delay (ADKD) type to define which navigation data are authenticated and which disclosed key should be used.
Despite the different ADKD types, tag generation follows the same algorithm:
\begin{equation}
    Tag = trunc_{TS}\left(MF(K, m_t)\right),
    \label{eq:tag}
\end{equation}
where \(MF\) is the configured MAC function, \(K\) is the delayed TESLA key, and \(trunc_{TS}(\cdot)\) truncates the output to the configured tag size. The message input can be expressed as
\begin{equation}
    m_t = PRN_D || PRN_A || GST_{sf} || CTR || navdata,
    \label{eq:mt}
\end{equation}
where \(PRN_D\) denotes the satellite whose data are authenticated, \(PRN_A\) denotes the satellite transmitting the OSNMA data, \(GST_{sf}\) is the subframe start time of Galileo System Time (GST), \(CTR\) is the tag position, and \(navdata\) contains the selected navigation data. If \(PRN_D=PRN_A\), the tag provides self-authentication; otherwise it provides cross-satellite authentication. 
Since E5b and E1 broadcast the same I/NAV messages at the subframe level, E1 OSNMA also supports cross-band authentication of E5b messages.

\subsection{SAS Basics}\label{sec:OSNMA Basics}
OSNMA is the foundation of SAS because its delayed keys are used to protect selected E6-C encrypted code fragments.
The public SAS workflow can be divided into two processes.
In the offline process, the Galileo service selects future fragments of the encrypted E6-C signal, denoted as Encrypted Code Sequences (ECS). 
These ECS fragments are re-encrypted with keys derived from future OSNMA TESLA keys, producing RECS files.
The receiver obtains RECS files, which contain RECS together with timing and processing metadata, via a secure server \cite{fernandez2023semiassisted,fernandez2024galileo}. 
A receiver downloads the files for the desired interval before or during operation.

\begin{figure}[h]
\centerline{\includegraphics[width=\linewidth]{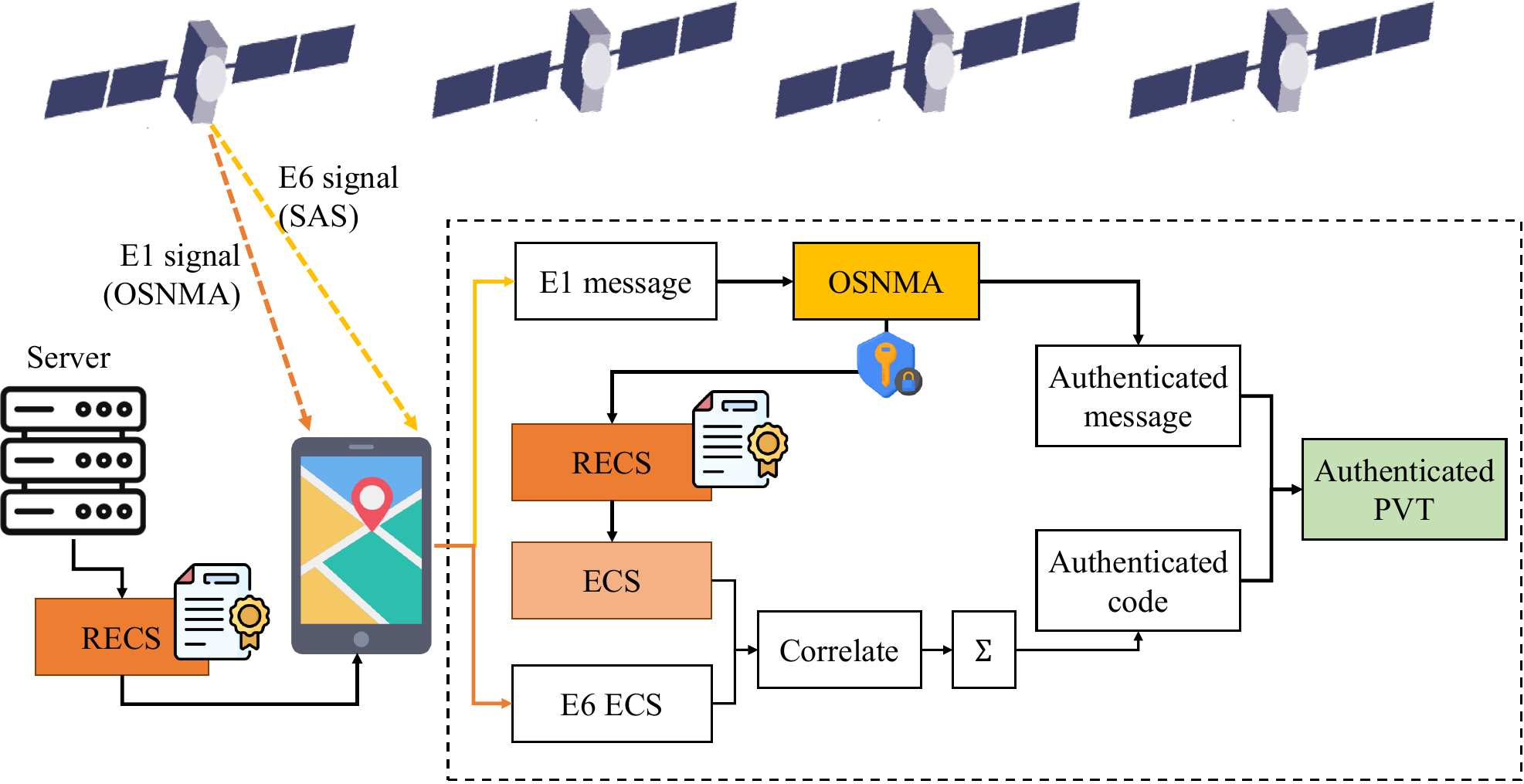}}
\caption{Galileo SAS operation.}
\label{fig:SAS-Overview}
\end{figure}

In the real-time process, Galileo satellites transmit E1 signal containing OSNMA data and the encrypted E6-C signal. 
When a RECS timestamp is reached, the receiver records an E6-C sample snapshot around the expected ECS epoch. 
After the OSNMA key is disclosed, the receiver derives the RECS decryption key, decrypts the RECS into a local ECS replica, and correlates it with the recorded ECS snapshot. 
If the correlation peak is present at a delay consistent with the E1-assisted timing, the receiver determines the corresponding E6-C measurement is authentic. 
These authenticated E6-C measurements can then be combined with OSNMA-authenticated navigation data by a receiver PVT engine, as shown in Fig.~\ref{fig:SAS-Overview}.

Several parameters in SAS directly affect satellite system and receiver behavior.
The RECS period determines the frequency of spreading code authentication.
The ECS length affects detection probability, storage, and processing cost. 
Randomization of RECS start times can reduce predictability but increases receiver scheduling complexity. 
These parameters prompted the development of a controllable simulator capable of implementing the SAS mechanism.

\subsection{Simulation Requirements}
The proposed SAS research platform meets the following requirements. 
First, it supports end-to-end signal simulation, generating signals that can be transmitted using SDR devices or processed offline. 
The receiver should be able to track the generated signals, parse navigation messages and pseudoranges, and perform authentication, rather than just isolated encryption tests. 
Second, it supports configurable parameters. Researchers can control satellite visibility, receiver target position, time, signal parameters, OSNMA parameters, and SAS parameters. 
Third, it must be reproducible. The same scenario should produce the same authentication events and receiver outputs, enabling controlled comparisons of different receiver designs. 
Fourth, it can publicly disclose intermediate states, such as visible satellite sets, generated tags, TESLA keys, RECS/ECS records, and measurements.

\section{Platform Design and Implementation}\label{sec:design}

\subsection{Design Overview}
\begin{figure}[h]
\centerline{\includegraphics[width=\linewidth]{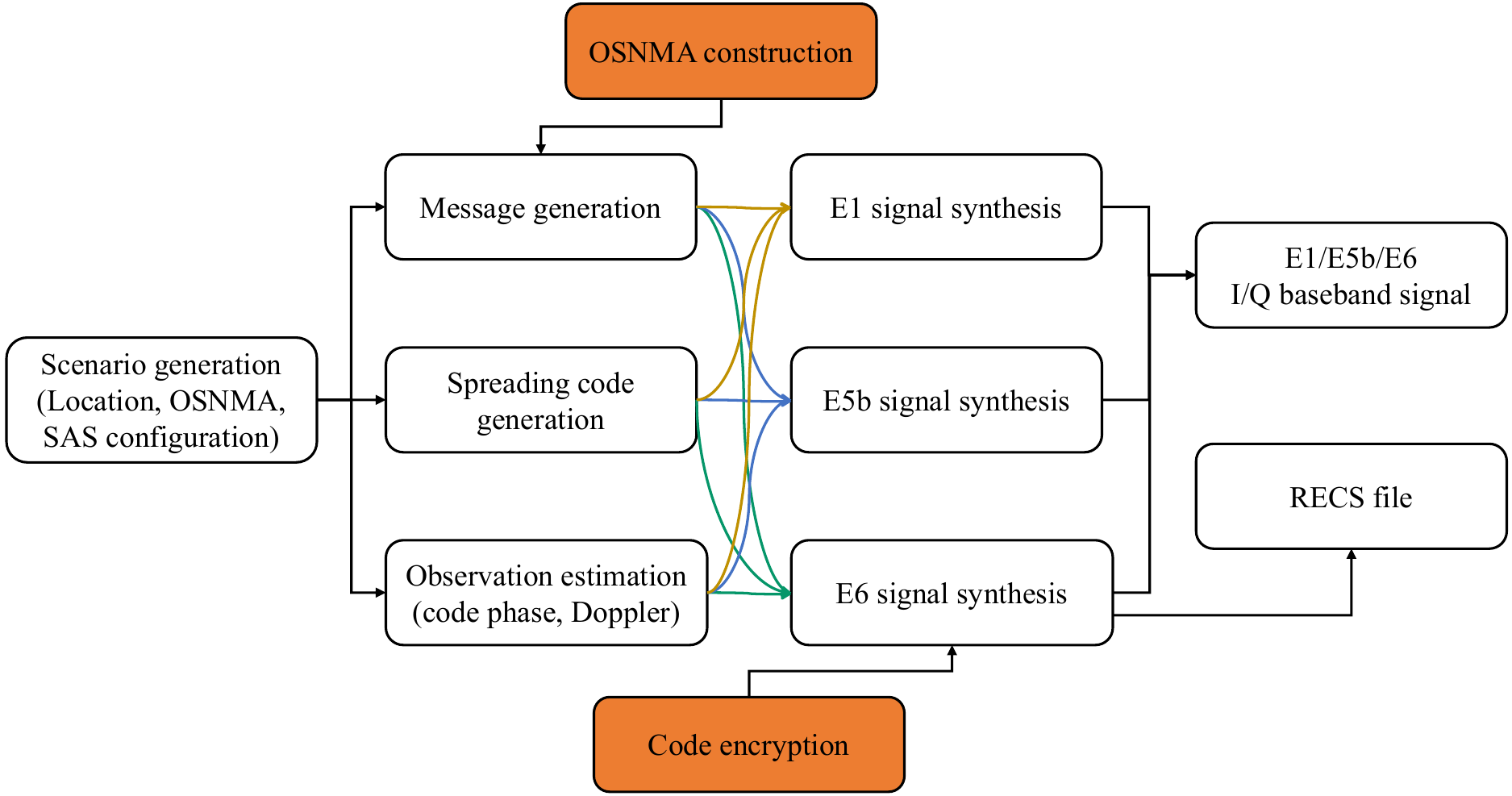}}
\caption{GalSAS-SDR-SIM platform architecture.}
\label{fig:GalSAS-SDR-SIM}
\end{figure}

GalSAS-SDR-SIM is built on a modular architecture, including modules for scenario generation, navigation message generation, OSNMA authentication data construction, spreading code generation and encryption, observation estimation,  baseband signal synthesis, and SDR transmission. 
Fig.~\ref{fig:GalSAS-SDR-SIM} illustrates the architecture of the GalSAS-SDR-SIM platform. 
The platform takes receiver location, start time, duration, ephemeris source, visible satellite strategy, OSNMA configuration, and SAS configuration as input. 
It then generates synchronized E1, E5b, and E6 baseband signals. The E1 signal carries I/NAV and OSNMA data, the E5b signal is used for cross-band authentication, and the E6 signal carries the research-oriented SCA component.
To verify the effectiveness of the platform, we further developed a prototype receiver, based on the open-source GNSS-SDR software \cite{gnsssdr38}, that supports OSNMA authentication and SAS authentication.

Signal generation follows the Galileo open service signal ICD \cite{galileosis}, while OSNMA follows the publicly available OSNMA ICD and receiver guidelines \cite{OSNMAreceiver21,OSNMAsisicd20}. 
The SCA components follow publicly available SAS documentation and supporting reproducible studies, such as encryption and scheduling of RECS/ECS and correlation-based verification \cite{fernandez2023semiassisted}.

\subsection{Unified Signal Generation}
The unified signal generator is the entry point of GalSAS-SDR-SIM. 
Its function is to generate the Galileo band signals required for OSNMA and SAS based on the user-specified scenario.
This is essential for SAS evaluation, because E1 OSNMA key disclosure, E6-C ECS transmission, and receiver-side correlation must refer to the same satellite geometry, Doppler, and code-phase evolution.

The generator first loads Galileo ephemeris from public RINEX files \cite{eph} and determines the visible satellites for a user-defined receiver position, start time, duration, and elevation mask \cite{galileo-sim}. 
It then constructs band-specific signal components. 
For E1, the platform generates E1-B I/NAV pages and inserts OSNMA fields.
For E5b, it generates I/NAV pages for cross-band authentication.
For E6, it generates the E6-B/E6-C spreading-code components needed for SCA experiments. 

For every visible satellite, the platform computes pseudorange, code phase, carrier Doppler, and update intervals from the same scenario \cite{understandGPS61,softwaredefindGPS62}. 
These values drive sample-level synthesis for all enabled bands, so the generated E1, E5b, and E6 IQ files remain mutually consistent. 
The output can be stored as deterministic baseband files for offline GNSS-SDR processing or streamed through SDR hardware for over-air experiments. 

\subsection{OSNMA Design} 
A complete OSNMA receiver check includes root key authentication, TESLA key chain authentication, and tag authentication. To keep the generated signal compatible with existing OSNMA receivers, GalSAS-SDR-SIM imports auxiliary OSNMA material, including signed root key information, TESLA key chain, and disclosure timing, from public test vectors or recorded Galileo data. 
The platform then regenerates authentication tags for the simulated navigation data while preserving the original root and key verification.

The OSNMA module supports three authentication modes. 
In self-authentication mode, a satellite authenticates its own E1 I/NAV data. 
In cross-band authentication, the platform strictly ensures consistency between E1 and E5b navigation messages, thus naturally supporting the use of E1 tags to authenticate E5b navigation messages. 
In cross-satellite authentication, satellites transmitting OSNMA data (called connected satellites) authenticate navigation data from satellites that do not transmit OSNMA (called disconnected satellites).
In addition to configurations with connected and disconnected satellite sets, the platform also supports cross-satellite authentication between connected satellites, or a fully connected configuration where all satellites transmit OSNMA data. This allows the platform to evaluate partial OSNMA deployments, reduced E1 availability, and authentication continuity.

The main design issue in cross-satellite authentication is tag allocation. 
Each MACK message has a limited number of fixed and flexible tag slots determined by MACLT. These slots determine not only which satellites are authenticated, but also when their authentication tag becomes available to the receiver. 
GalSAS-SDR-SIM therefore implements a configurable tag scheduler. 
Users can select the connected satellites, choose the number of disconnected satellites assigned to each connected satellite, and evaluate different allocation policies.

\begin{figure}[h]
\centerline{\includegraphics[width=\linewidth]{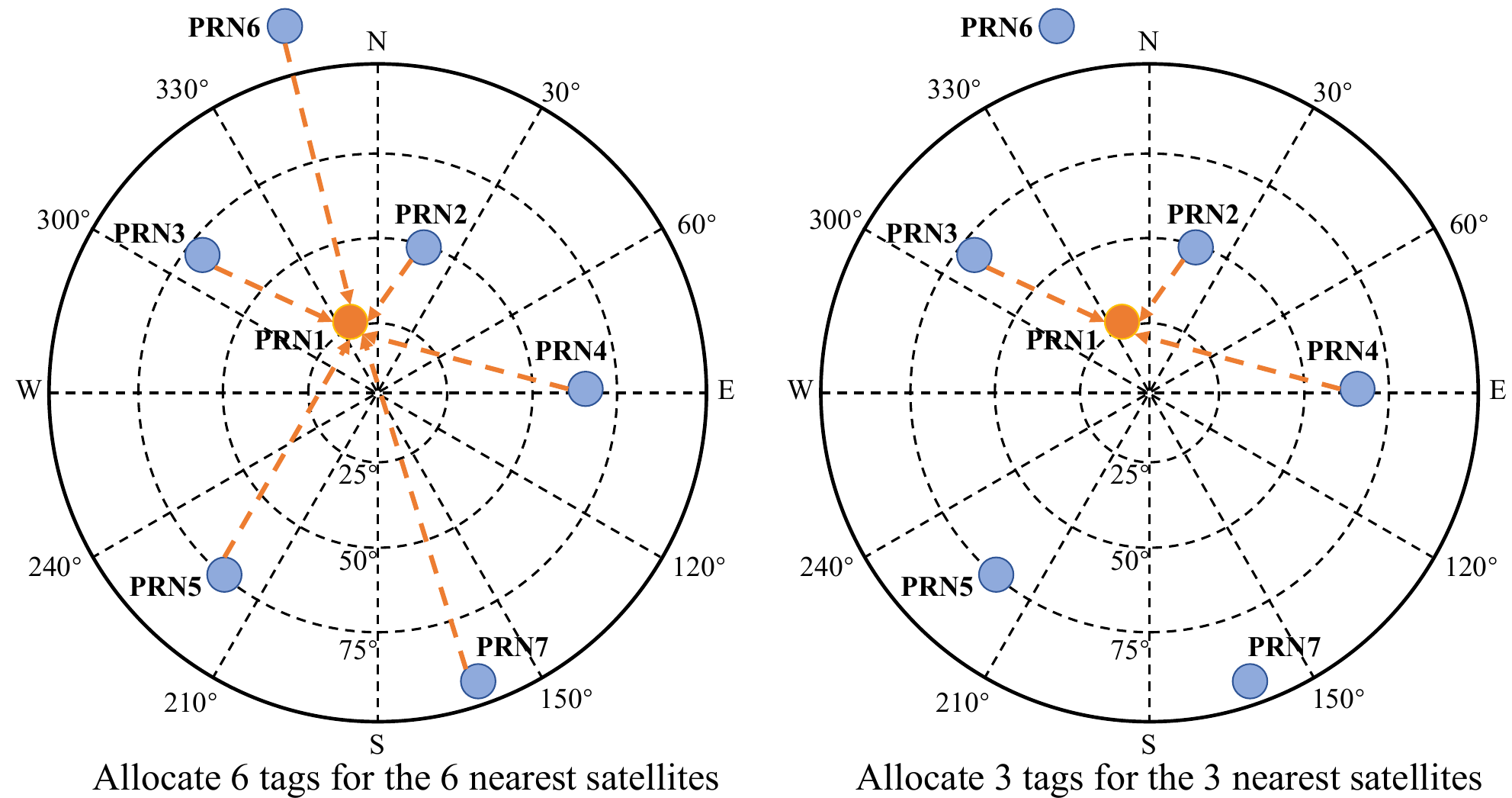}}
\caption{Cross-satellite authentication under different parameters \(N_{ca}\).}
\label{fig:cross-sat}
\end{figure}

The current scheduler uses a geometry-aware policy inspired by OSNMA cross-authentication analysis \cite{galan2025sensitivity}. 
For each connected satellite, the platform ranks candidate disconnected satellites by inter-satellite distance and assigns the closest \(N_{ca}\) candidates to available cross-authentication slots. 
A larger \(N_{ca}\) increases coverage and redundancy, while a smaller \(N_{ca}\) concentrates tags on satellites more likely to be jointly visible. 
Fig.~\ref{fig:cross-sat} illustrates two allocation densities. 
This module gives the later evaluation a direct way to measure authenticated-satellite availability, time to first authenticated fix, and robustness under partial OSNMA deployment.

\begin{figure}[h]
\centerline{\includegraphics[width=\linewidth]{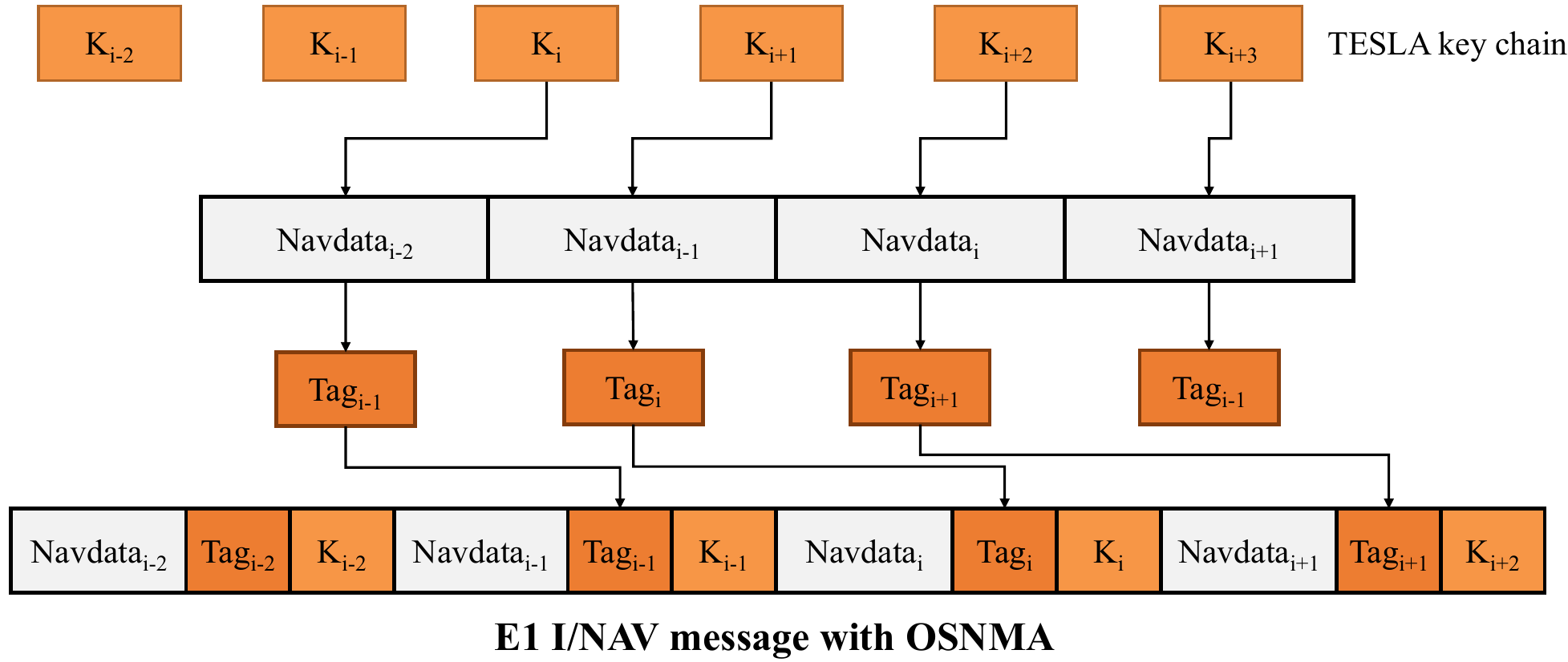}}
\caption{OSNMA generation process.}
\label{fig:osnma-workflow}
\end{figure}

Once the tag allocation strategy is determined, the platform generates corresponding tags for each subframe of all visible satellites based on Eq.~\ref{eq:tag}.
The platform embeds these tags into the corresponding position in the constructed OSNMA field based on the declaration of the MACLT parameters. 
Then, it embeds the reused root key and TESLA key into the constructed OSNMA field. 
Finally, the OSNMA field is embedded into the navigation message to support all OSNMA authentication steps, as shown in Fig.~\ref{fig:osnma-workflow}.

\subsection{SAS Design}
The SAS module is the core of the GalSAS-SDR-SIM platform. 
Public SAS studies describe a semi-assisted architecture, i.e., the system encrypts the Galileo E6-C signal, selects short encrypted fragments as ECS, re-encrypts these fragments as RECS using future OSNMA TESLA keys, and lets the receiver authenticate the stored E6-C snapshots after the corresponding E1 key is disclosed \cite{fernandez2023semiassisted,fernandez2024galileo,fernandez2026prototyping}. 
GalSAS-SDR-SIM follows this logic, but implements it as a configurable SDR experiment platform rather than as a fixed service emulator.

\begin{figure}[h]
\centerline{\includegraphics[width=\linewidth]{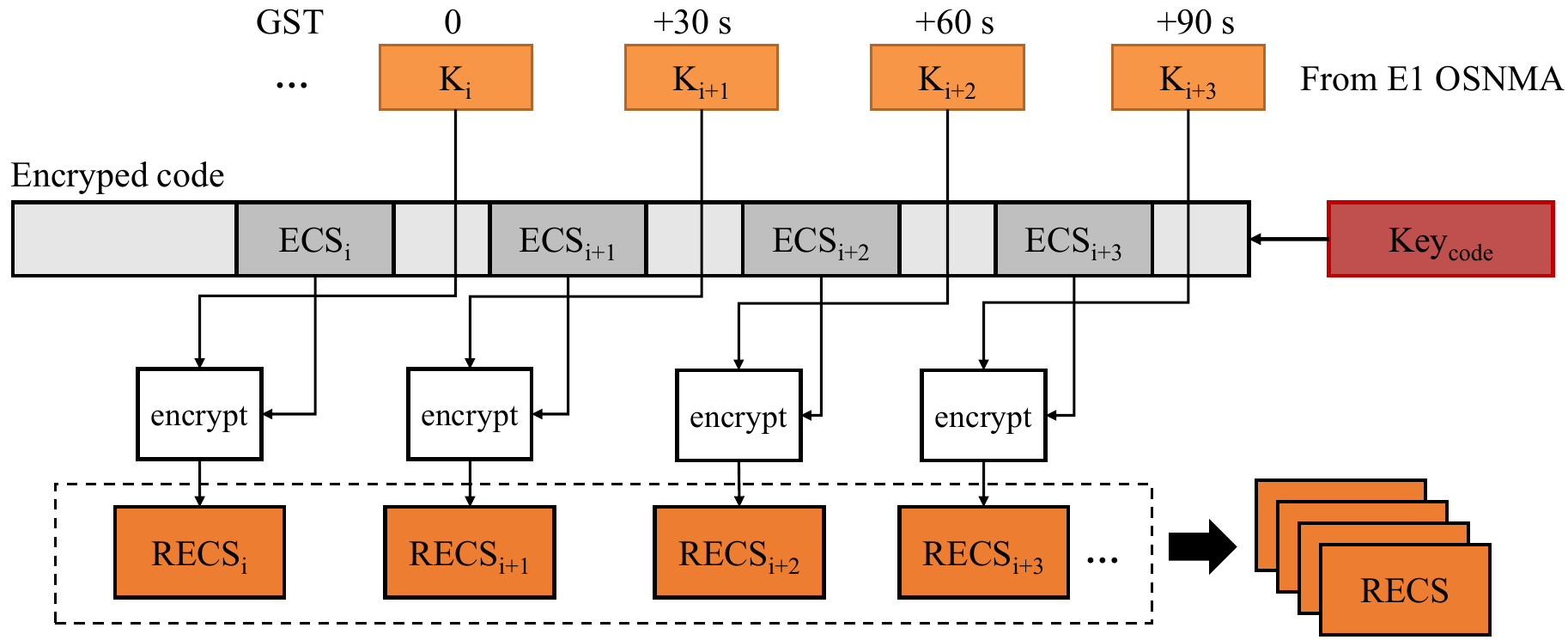}}
\caption{SAS generation process.}
\label{fig:sas-workflow}
\end{figure}

The SAS generation consists of two stages: E6-C encryption construction and ECS/RECS generation, as shown in Fig.~\ref{fig:sas-workflow}. 
The first stage produces a continuous encrypted E6-C code stream. 
A configurable 256-bit master key is expanded into satellite-specific keys, and an AES-based stream is used to flip the E6-C codes for the selected satellites. 
This stage provides the unpredictable signal component required by SCA experiments.

The second stage builds RECS records from the encrypted stream. 
For each authenticated satellite, the platform divides time into RECS periods and selects one ECS window inside each period. 
The selected ECS is re-encrypted with a key derived from the same OSNMA TESLA key schedule embedded in the generated E1 signal. 
In the current implementation, the delayed key setting is fixed to one TESLA disclosure interval in the evaluation, so the receiver can authenticate an earlier E6-C fragment only after the corresponding E1 key has been verified. 
This design keeps E1 message authentication and E6-C code authentication temporally and cryptographically coupled.

\begin{table}[H]
\centering
\caption{Configurable SAS configuration fields exposed by GalSAS-SDR-SIM.}
\label{tab:sas-configuration-fields}
\begin{tabularx}{\linewidth}{lX}
\hline
Field & Function in the platform \\
\hline
RECS period & Controls the frequency of authentication generation; supported values include 100, 200, 500, and 1000 ms. \\
ECS length & Controls the duration of the encrypted code fragment; supported values include 4, 8, and 16 ms. \\
Window position & Places the ECS at the beginning, second 16-ms interval, or end of the RECS period. \\
Configuration ID & Identifies the active configuration and binds records to a specific experiment configuration. \\
Randomization and noise & Support repeatable randomized windows and deterministic Additive white Gaussian noise (AWGN) experiments. \\
\hline
\end{tabularx}
\end{table}

Following public SAS terminology, we distinguish ECS, RECS, RECS records, and RECS files.
ECS denotes the encrypted E6-C code fragment selected from the signal timeline. 
RECS denotes the re-encrypted code sequence protected by an OSNMA-derived key.
A RECS record refers to one such authentication object for a specific satellite, ECS epoch, and key. 
In GalSAS-SDR-SIM, the receiver reads RECS files that store multiple RECS records together with the metadata needed for snapshot scheduling and ECS reconstruction. 
Thus, RECS is the authentication content, whereas a RECS file is a data product provided by the platform to the receiver.

The RECS file makes the same SAS configuration usable by the generator and the receiver. 
Its header stores the sampling rate, RECS period, ECS length, window position, key timing, chip count, and payload size. 
Each record stores the satellite identifier, ECS epoch, TESLA key tag, initialization vector (IV), and encrypted ECS payload. 
The IV is derived from the authenticated GST time and record context using SHA-256, avoiding IV reuse when multiple satellites or multiple ECS windows share the same disclosed TESLA key. 

The receiver design is also configuration-driven. 
The modified GNSS-SDR receiver reads the SAS configuration from the RECS file rather than duplicating these parameters in a separate receiver configuration. 
It then dynamically sets the expected ECS chip count, ciphertext length, snapshot duration, padding check, replica construction, and correlation workload. 
In normal E1-assisted mode, the receiver is not allowed to read a local key chain file. 
It must obtain TESLA keys from the received E1 signal after OSNMA verification, including root key signature verification and TESLA chain validation.  
Inconsistent keys, times, IVs, record headers, or windows cause the receiver to fail closed and reject SAS authentication.

Because the continuous E1/E6 IQ stream and the SAS configuration are decoupled, different configurations can be evaluated using the same underlying signal scenario while regenerating only the RECS file. 
Shorter RECS periods increase authentication freshness but increase file size, snapshot scheduling, and receiver workload. 
Longer ECS windows provide more samples for correlation, especially under noise, but increase storage and processing cost. 
Window placement affects snapshot timing and range-bias sensitivity. 
The exported CSV and summary JSON report PASS/WARMUP/FAIL/ERROR counts, effective authentication rate, correct-key and wrong-key correlation peaks, processing time, pending storage, authentication latency, and range-bias statistics. 
The configurable parameters supported by the platform are shown in Table~\ref{tab:sas-configuration-fields}, which provides the basis for the evaluation in Section~\ref{sec:evaluation}.

\section{Experiment and Evaluation} \label{sec:evaluation}
This evaluation has two goals. 
First, we verify that GalSAS-SDR-SIM can generate receiver-verifiable OSNMA data, because SAS relies on the E1 key-disclosure timeline. 
Second, we evaluate the SAS workflow itself, including end-to-end E1-assisted E6-C authentication, configuration sensitivity, noise robustness, and receiver resource cost.
The authenticated PVT calculation depends on the receiver-specific strategy and is not the focus of this evaluation.

\begin{figure}[h]
\centerline{\includegraphics[width=\linewidth]{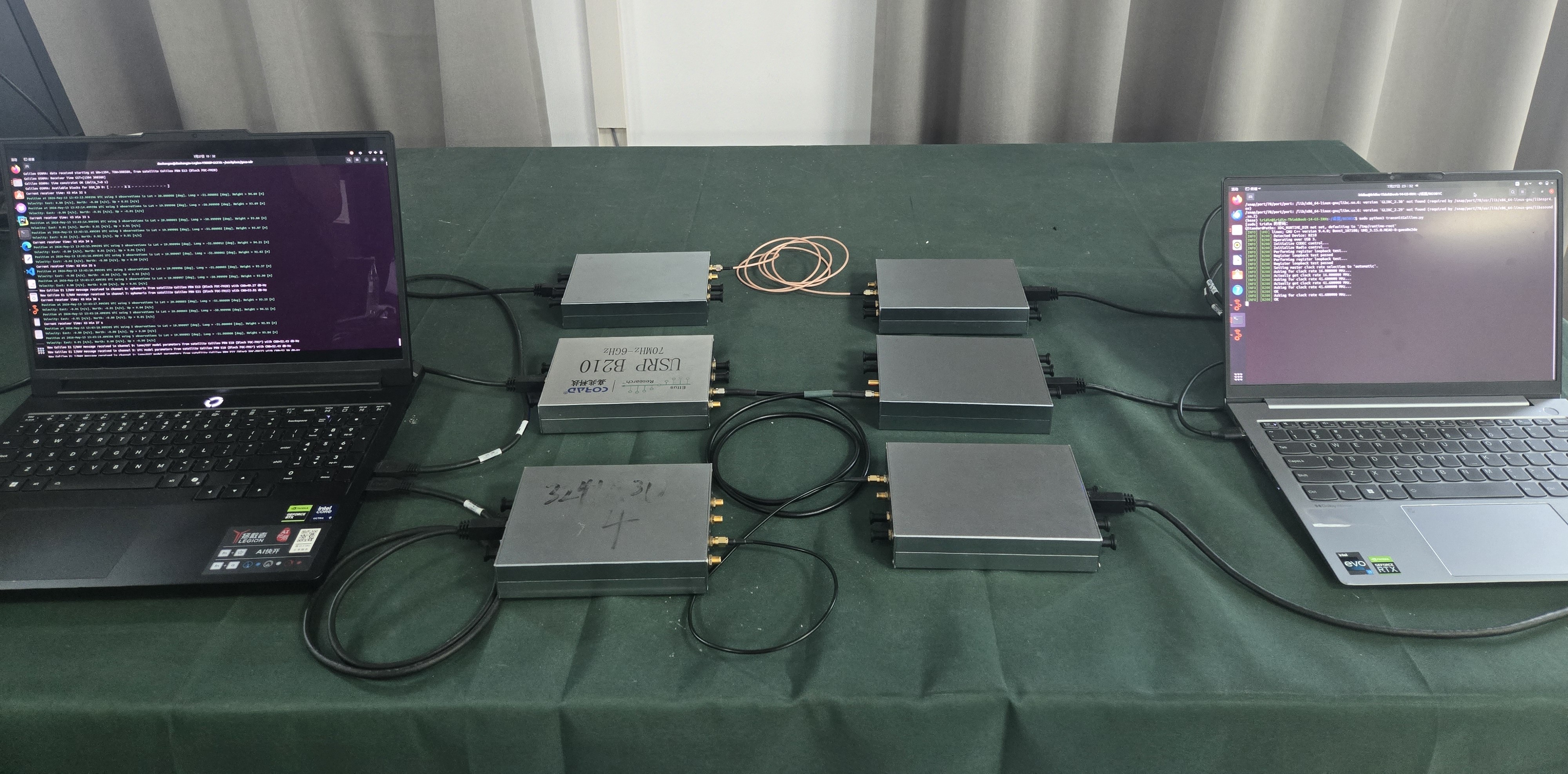}}
\caption{Experiment setup.}
\label{fig:Experiment-setup}
\end{figure}

The experimental setup is shown in Fig.~\ref{fig:Experiment-setup}. We use three USRP B210s to transmit E1, E5b and E6 signals at their respective frequencies, and use another three USRP B210s to receive these signals.
Two auxiliary OSNMA datasets are used for evaluation: ESA public test vectors and self-collected datasets extracted from genuine Galileo signals on the rooftop of our laboratory. 
These datasets cover different OSNMA configurations, including different TESLA key chains. 
The evaluation metrics are shown in Table~\ref{tab:evaluation-metrics}.

The simulator generates synchronized Galileo E1, E5b and E6 baseband files from the same receiver location, start time, ephemeris, satellite visibility, pseudorange, Doppler, and code-phase model. 
E1 carries I/NAV and OSNMA, while E6 carries the encrypted E6-C component and the selected ECS windows. 
The receiver is the modified GNSS-SDR described in Section~\ref{sec:design}. 
In normal E1-assisted mode, the receiver is configured with E1 samples, E6 samples, RECS files, and the public OSNMA verification material.

\begin{table}[h]
\centering
\caption{Main evaluation metrics.}
\label{tab:evaluation-metrics}
\begin{tabularx}{\linewidth}{lX}
\hline
Metric & Meaning \\
\hline
PASS/WARMUP/FAIL/ERROR & Per-record authentication status. WARMUP denotes records observed before the required OSNMA trust chain and disclosed key are available. \\
TTFAF & Time to first authenticated fix obtained using OSNMA \cite{galan2025TTFAF}. \\
Authenticated-satellite ratio & Fraction of time with at least four authenticated satellites available for positioning obtained using OSNMA \cite{hammarberg2024experimental}. \\
Correlation margin & Separation between correct-key and wrong-key E6-C correlation peaks. \\
Range bias & Difference between the E1-assisted expected delay and the accepted E6-C correlation delay. \\
Latency & Time between ECS observation and availability of the corresponding authenticated result. \\
Resource cost & RECS file size, pending snapshot storage, and receiver processing time. \\
\hline
\end{tabularx}
\end{table}

\subsection{OSNMA Evaluation as the Authentication Backbone}
Although the paper focuses on SAS, OSNMA must be evaluated because it provides the navigation message authentication and the delayed keys used for RECS decryption. 
Therefore, we evaluate the performance of OSNMA authentication, focusing primarily on tag authentication and the visibility of connected satellites.
As a representative case, the target position is set to (\(20^\circ\)N, \(-51^\circ\)E, 100 m), while the target start time is set to 2026-05-13 12:11:01 UTC. 

\begin{table}[t] \centering \caption{Tag processing and authentication results.} \label{tab:tag_authentication_results} \begin{tabular}{lrrrr} \toprule \textbf{Tag type} & \textbf{Sim.} & \textbf{Processed} & \textbf{Authen.} & \textbf{Success rate} \\ \midrule Total tags & 25191 & 18028 & 17951 & 99.57\% \\ Self tag & 10797 & 10504 & 10437 & 99.36\% \\ Cross tag & 14394 & 7524 & 7514 & 99.87\% \\ ADKD=0 & 16789 & 10391 & 10367 & 99.77\% \\ ADKD=4 & 2097 & 2078 & 2074 & 99.81\% \\ ADKD=12 & 6305 & 5559 & 5510 & 99.12\% \\ ADKD=0(S) & 4224 & 4107 & 4091 & 99.61\% \\ ADKD=4(S) & 2097 & 2078 & 2074 & 99.81\% \\ ADKD=12(S) & 4476 & 4319 & 4272 & 98.91\% \\ ADKD=0(C) & 12565 & 6284 & 6276 & 99.87\% \\ ADKD=12(C) & 1829 & 1240 & 1238 & 99.84\% \\ \bottomrule \end{tabular} \end{table}

\begin{figure}[h]
\centerline{\includegraphics[width=\linewidth]{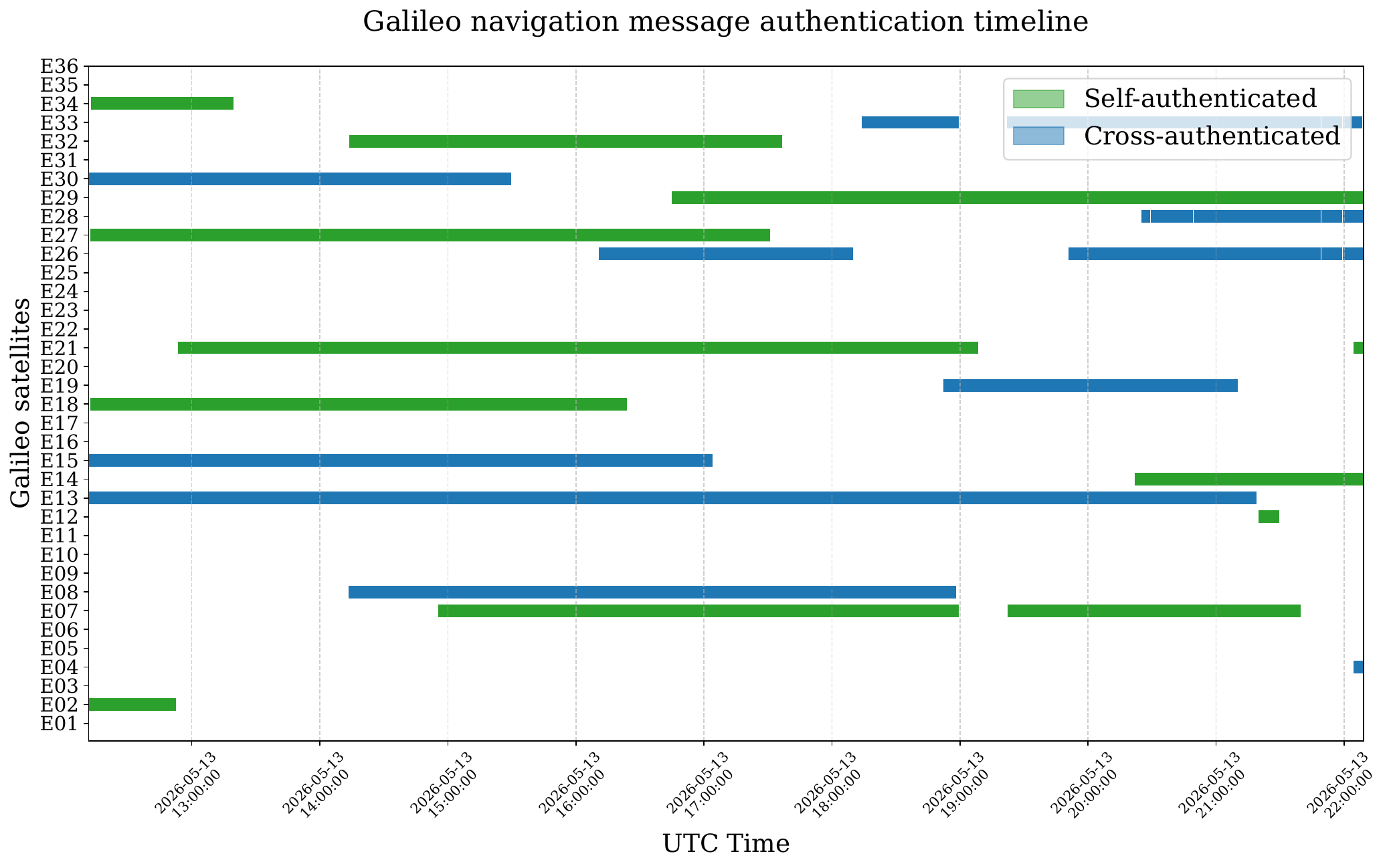}}
\caption{OSNMA authentication timeline under 10 hours of simulation.}
\label{fig:authen-timeline}
\end{figure}

Table~\ref{tab:tag_authentication_results} summarizes the number of tags generated during the 10-hour simulation and the number processed and authenticated by GNSS-SDR. 
In total, the platform generates 25,191 tags, of which 18,028 are processed by the receiver and 17,951 are authenticated, corresponding to an overall success rate of 99.57\%. 
Because some cross-authentication tags originated from satellites outside the field of view, the navigation data associated with these tags cannot be retrieved, resulting in a gap in the number of simulated and processed tags. 
Fig.~\ref{fig:authen-timeline} shows the authentication status timeline for visible satellites over a 10-hour period. At the beginning of the simulation, PRN 13, 15, and 30 do not transmit OSNMA data, but their tags are transmitted by PRN 2, 18, 27, and 34, so these satellites can still be authenticated through cross-satellite authentication.

\begin{figure}[h]
\centerline{\includegraphics[width=\linewidth]{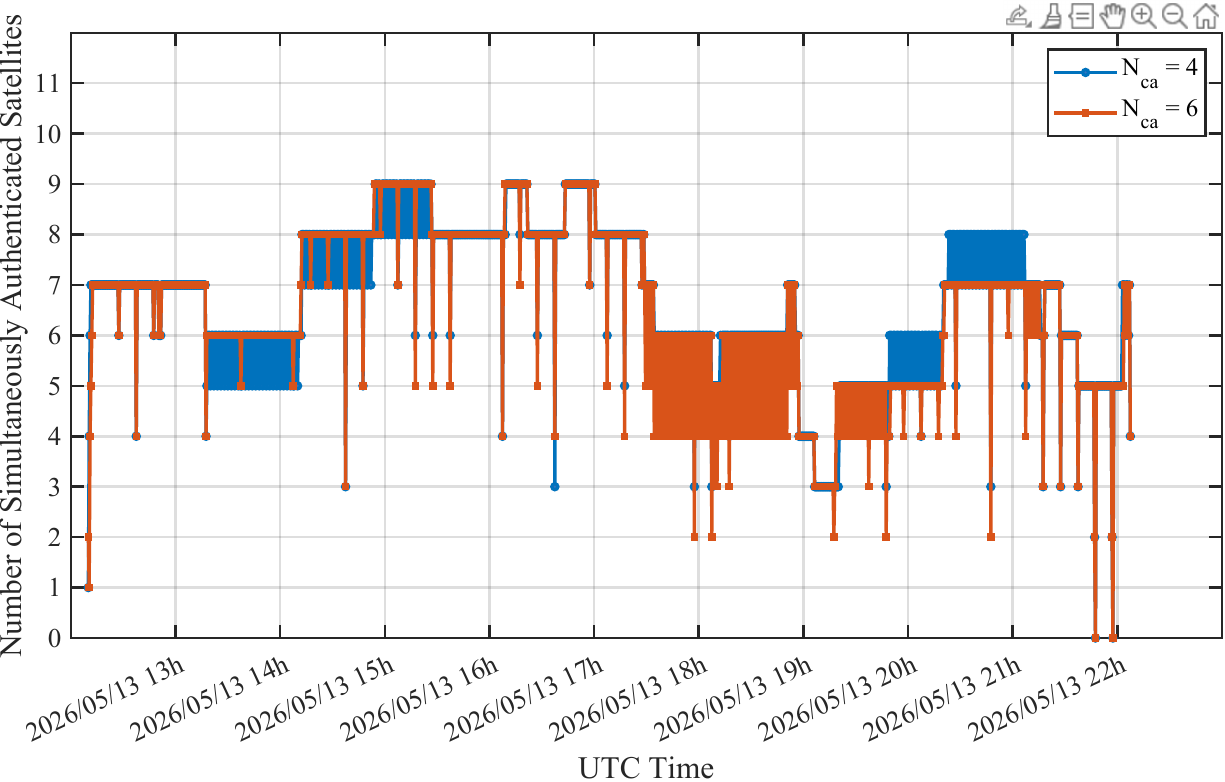}}
\caption{Number of simultaneously authenticated satellites over time.}
\label{fig:Num-authen-sat}
\end{figure}

We then count the number of simultaneously authenticated satellites every 30 seconds, which is important for continuous positioning. Fig.~\ref{fig:Num-authen-sat} compares the strategies of \(N_{ca}=4\) and \(N_{ca}=6\). 
In both strategies, the 95th percentile is 9 satellites. The time ratio with at least four authenticated satellites (i.e., Authenticated-satellite ratio) is 96.49\% and 96.24\%, respectively, and the average numbers of authenticated satellites are 6.69 and 6.47.

\begin{figure}[h]
\centerline{\includegraphics[width=\linewidth]{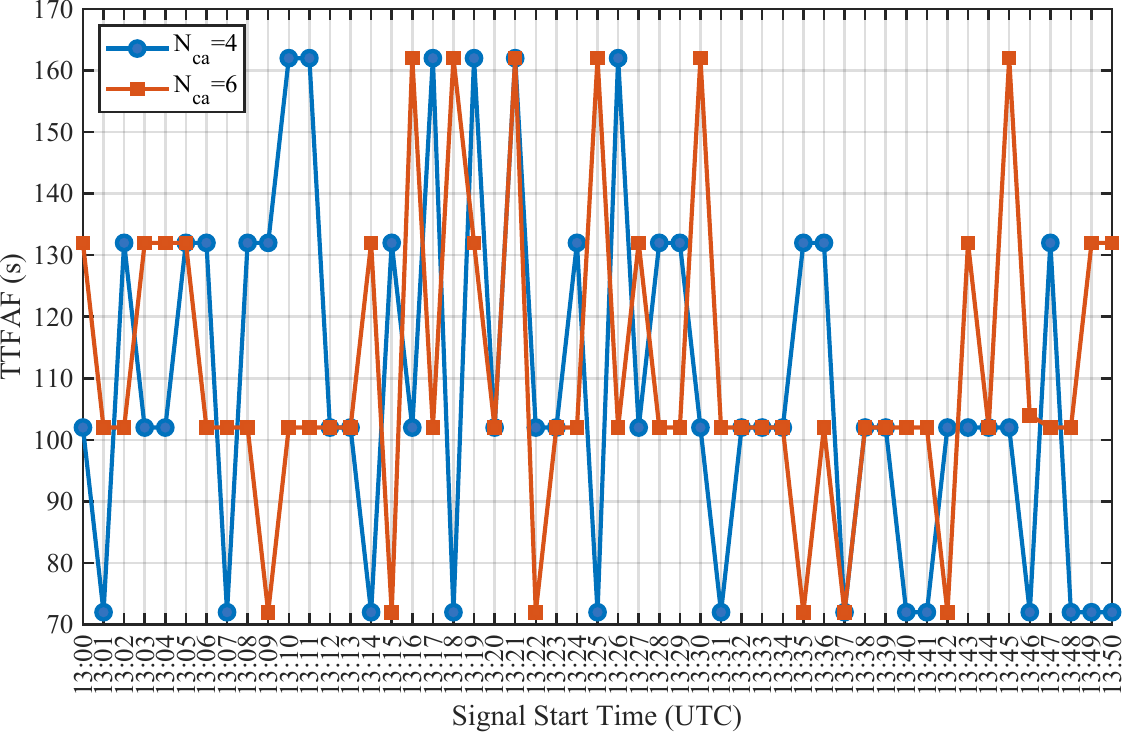}}
\caption{TTFAF under different \(N_{ca}\) and signal start time.}
\label{fig:TTFAF}
\end{figure}

We further evaluate the time to first authenticated fix (TTFAF) \cite{galan2025TTFAF}. 
Since a GNSS receiver may be powered on at an arbitrary time, we generate signal segments from 13:00:00 UTC to 13:50:00 UTC on 2026-05-13, with a start-time interval of 1 minute. Each segment is processed independently to emulate a different receiver startup time. Fig.~\ref{fig:TTFAF} presents TTFAF results from 102 independent experiments. For \(N_{ca}=4\), the mean and median TTFAF are 108.47 s and 102 s. For \(N_{ca}=6\), the mean increases slightly to 111.45 s, while the median remains 102 s. 
In both strategies, the minimum and maximum TTFAF values are 72 s and 162 s.

The platform also utilizes OSNMA authentication in E1 to authenticate navigation messages in E5b. 
Because the navigation messages and tags are matched, this verifies the simulator's cross-band authentication path.
We do not report a separate E5b performance study because SAS is the focus of this paper.
These results show that the OSNMA module can generate receiver-verifiable authentication data and provide the key-disclosure timeline required by the SAS experiments.

\subsection{End-to-End SAS Validation}
The baseline SAS experiment validates the complete path from synchronized signal generation to receiver-side authentication. 
The simulator generates E1 OSNMA and encrypted E6-C signals in the same scenario. 
It also produces an RECS file whose RECS records are protected using the same TESLA key schedule carried by E1. 
The receiver must first acquire and track E1, verify OSNMA root and TESLA-chain information, obtain disclosed keys from the received E1 signal, decrypt RECS records, reconstruct ECS replicas, correlate them with stored E6-C snapshots, and apply the range-gate and wrong-key checks.

The representative baseline configuration uses a 200-ms RECS period, a 16-ms ECS length, the second-window position, and a one-interval key delay. 
In a 600-s run, the receiver produced 4050 PASS records and 4395 WARMUP\_OSNMA records, with 0 FAIL and 0 ERROR records, as summarized in Table~\ref{tab:sas-baseline}. 
The WARMUP records correspond to the receiver startup interval before the OSNMA trust chain and the required disclosed keys became available. 
After warm-up, all 4050 measured records passed the authentication test, with a wrong-key Probability of False Alarm ($P_{fa}$) of 0 and a mean correct-key correlation peak of 0.317 compared with 0.004 for the wrong-key control. 
This result indicates that the generated E1 key timeline, E6-C encryption, RECS file metadata, snapshot scheduling, and correlation search are mutually consistent.

\begin{table}[t]
\centering
\caption{Representative end-to-end SAS validation.}
\label{tab:sas-baseline}
\begin{tabularx}{\linewidth}{lX}
\hline
Item & Result \\
\hline
SAS configuration & 200-ms RECS period, 16-ms ECS, second-window ECS placement \\
Receiver status counts & 4050 PASS, 4395 WARMUP\_OSNMA, 0 FAIL, 0 ERROR \\
Detection result & Probability of Detection ($P_d$) = 100\%, wrong-key $P_{fa}$ = 0\% \\
Mean correlation peaks & Correct key: 0.317; wrong key: 0.004 \\
Latency & Average 45.1 s; minimum 30.2 s; maximum 60.0 s \\
Range-bias metric & RMSE 11.71 m; this metric is code-delay bias \\
\hline
\end{tabularx}
\end{table}

The average authentication latency is about 45.1 s, ranges from 30.2 to 60.0 s. 
This range is expected because SAS authentication waits for delayed OSNMA key disclosure rather than accepting an E6-C snapshot immediately. 
The effective authentication rate is 15 records/s for the target satellites under the 200-ms configuration, while the measured processing throughput is about 9.9 records/s on the evaluated receiver implementation. 
This result shows that the platform can evaluate both authentication correctness and the timing cost imposed by OSNMA key disclosure.

\subsection{SAS Configuration Sensitivity}
We next evaluate how SAS configuration affects authentication frequency, storage, pending snapshot memory, and per-record processing cost. 

\begin{figure}[h]
\centerline{\includegraphics[width=\linewidth]{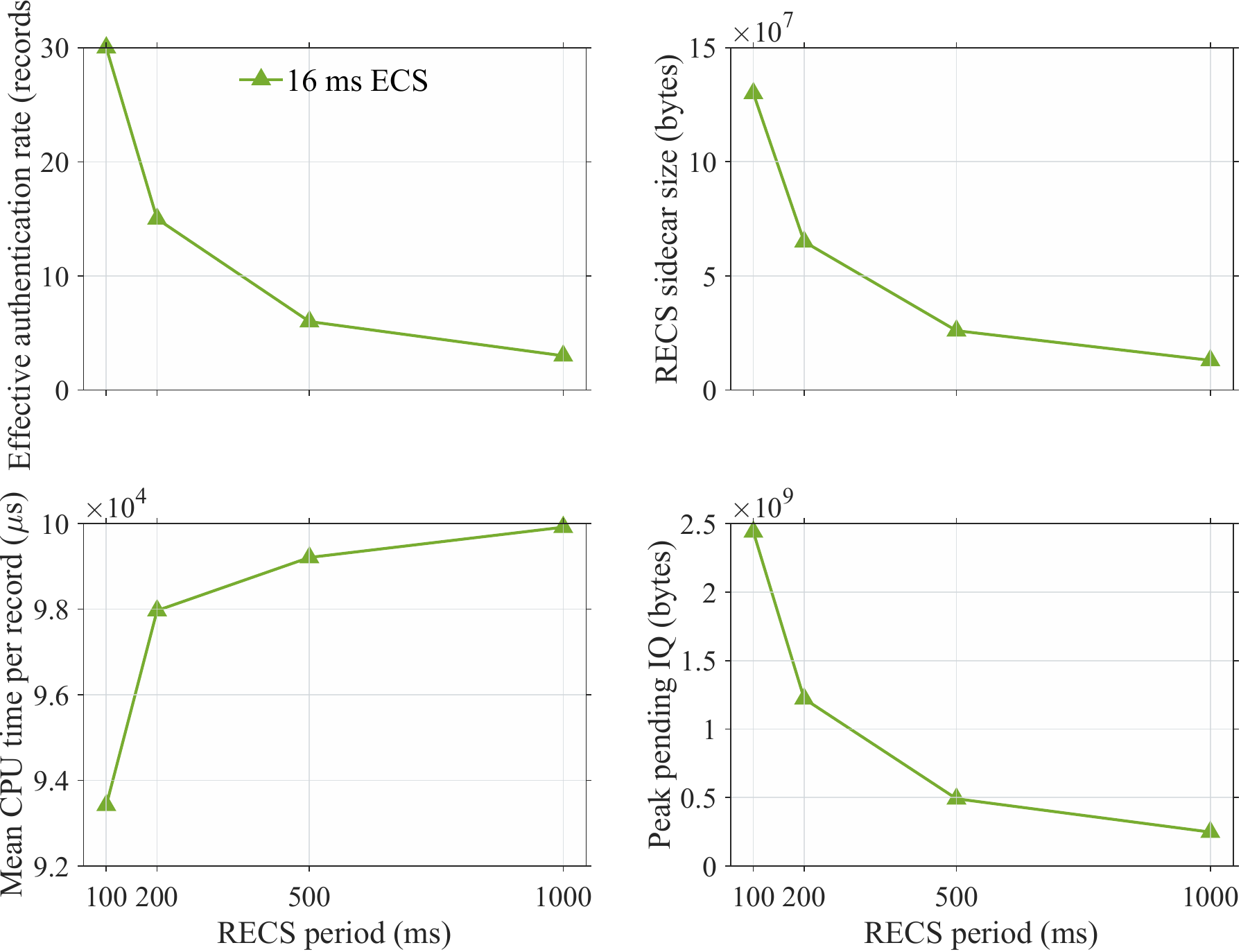}}
\caption{Trade-offs introduced by the RECS period under clean conditions.}
\label{fig:recs-period-tradeoffs}
\end{figure}

Fig.~\ref{fig:recs-period-tradeoffs} shows the RECS-period trade-off using 16-ms ECS windows under clean (i.e., no noise) conditions. 
Reducing the period from 1000 ms to 100 ms increases the effective authentication rate from 3 records/s to 30 records/s. 
The cost is proportional growth in the receiver-facing data volume and pending snapshot cache: the RECS file grows from 12.4 MB to 124.0 MB, and the peak pending IQ storage grows from 236.0 MB to 2326.5 MB. 
The per-record processing time is dominated by correlation and ranges from 93 to 100 ms for 16-ms ECS windows. 
Therefore, the RECS period mainly controls authentication density and buffering pressure, while the ECS length controls per-record computation.

\begin{figure}[h]
\centerline{\includegraphics[width=\linewidth]{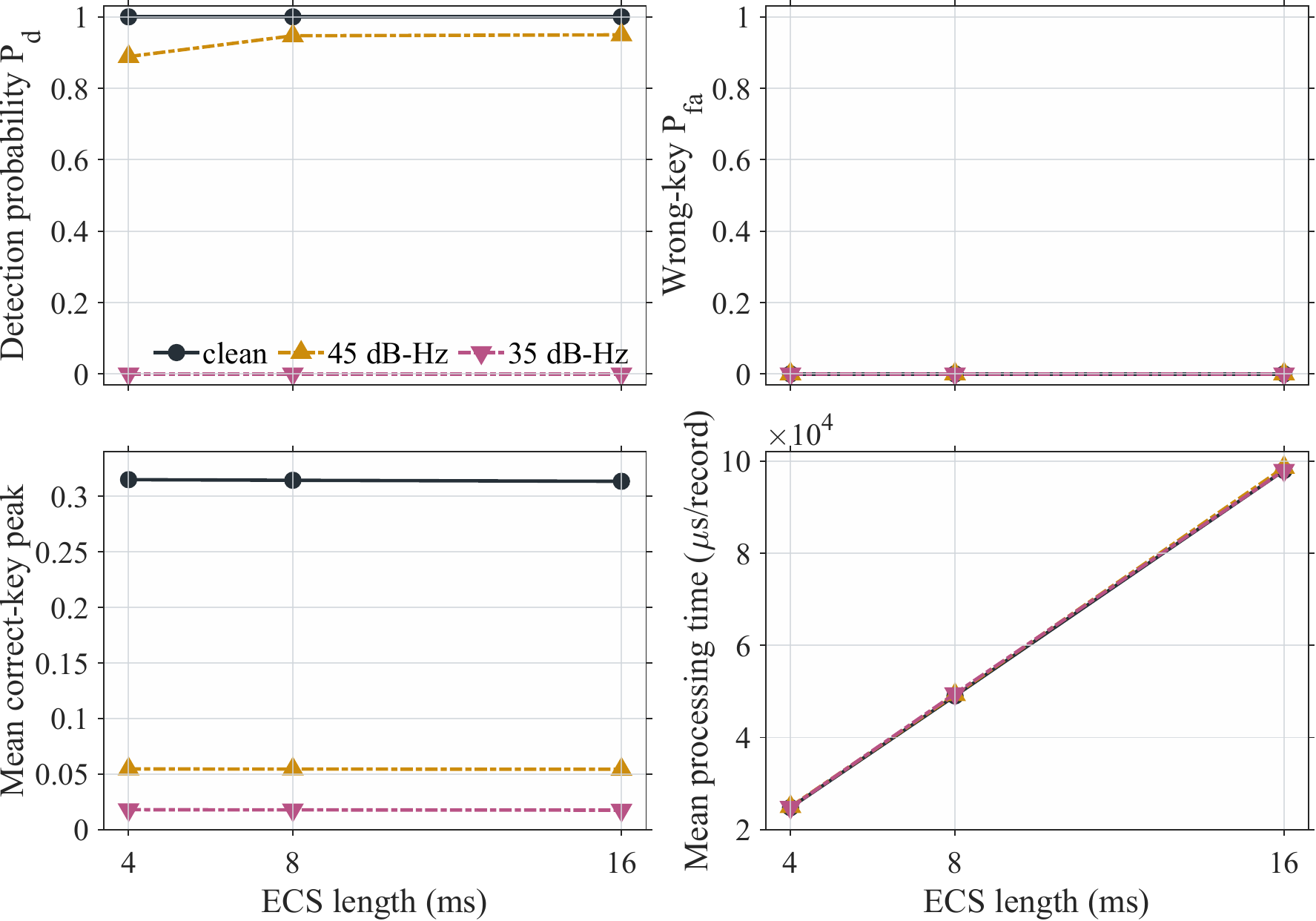}}
\caption{Detection and processing trade-offs introduced by the ECS length.}
\label{fig:ecs-length-tradeoffs}
\end{figure}

Fig.~\ref{fig:ecs-length-tradeoffs} compares ECS lengths of 4, 8, and 16 ms under a fixed 200-ms RECS period and second-window placement. 
Under clean conditions all three ECS lengths achieve 100\% PASS and 0 wrong-key false alarms. 
However, the receiver workload scales strongly with the ECS length. 
The average total processing time increases from 24.8 ms for 4-ms ECS to 49.0 ms for 8-ms ECS and 98.0 ms for 16-ms ECS. 
The RECS file and pending storage show the same trend, increasing from 15.8 MB and 291.3 MB for 4-ms ECS to 62.0 MB and 1165.1 MB for 16-ms ECS.
This indicates that shorter ECS windows are more efficient, while longer ECS windows can provide more samples for robust detection under noise.

All measured records pass under all configurations, which verifies that the RECS file generator and receiver parser remain consistent across different RECS periods, ECS lengths, and window positions. 
The range-bias root mean square error (RMSE) remains within a narrow interval of 11.70--12.04 m across these configurations, indicating that changing the SAS configuration does not introduce a systematic code-delay error in clean conditions. 
The most visible difference is resource cost.
Setting ECS windows of 4-ms and 8-ms reduces the processing time and pending storage, while 100-ms RECS periods increase the number of authentication records and buffering demand.


\begin{figure}[h]
\centerline{\includegraphics[width=\linewidth]{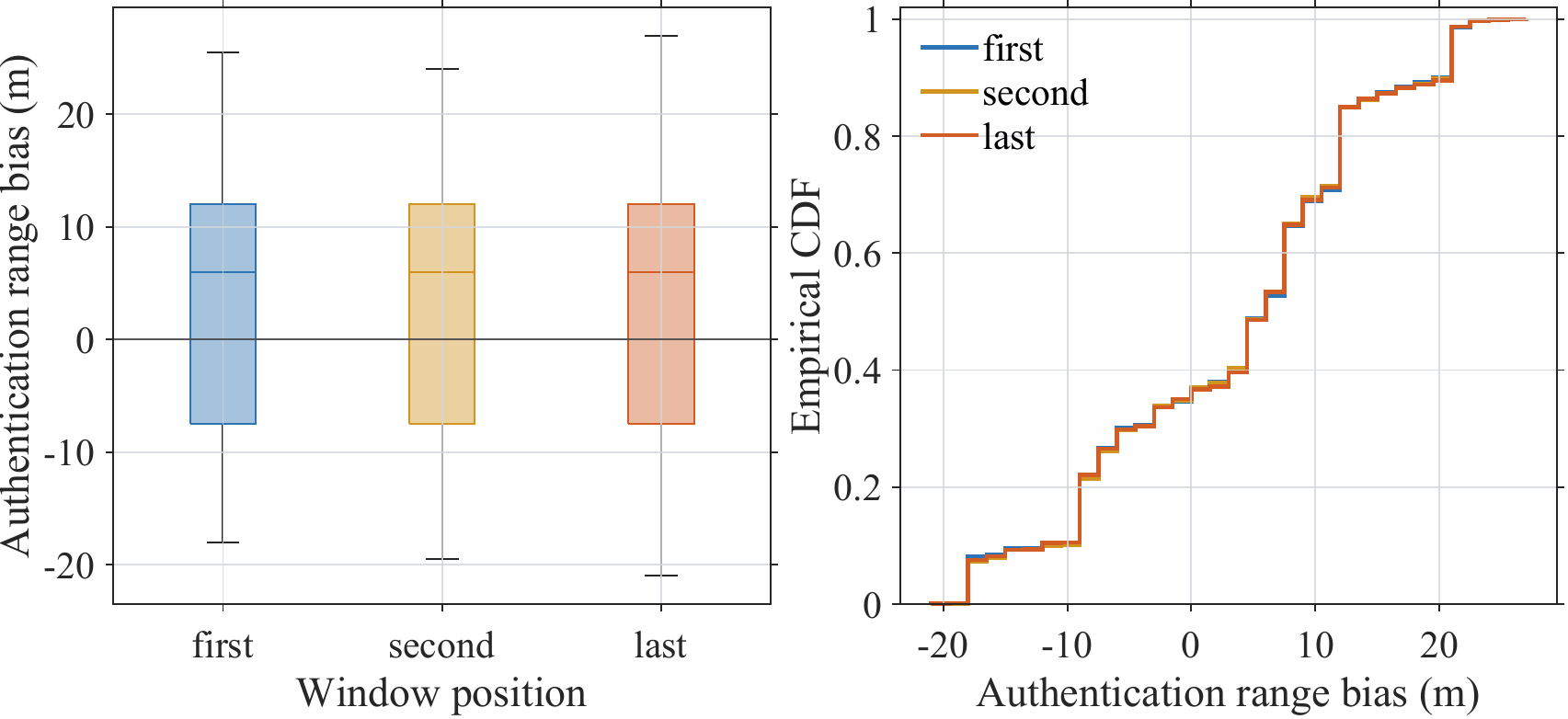}}
\caption{Authentication range-bias distribution under different ECS window positions.}
\label{fig:window-position-range-bias}
\end{figure}

\begin{figure}[h]
\centerline{\includegraphics[width=\linewidth]{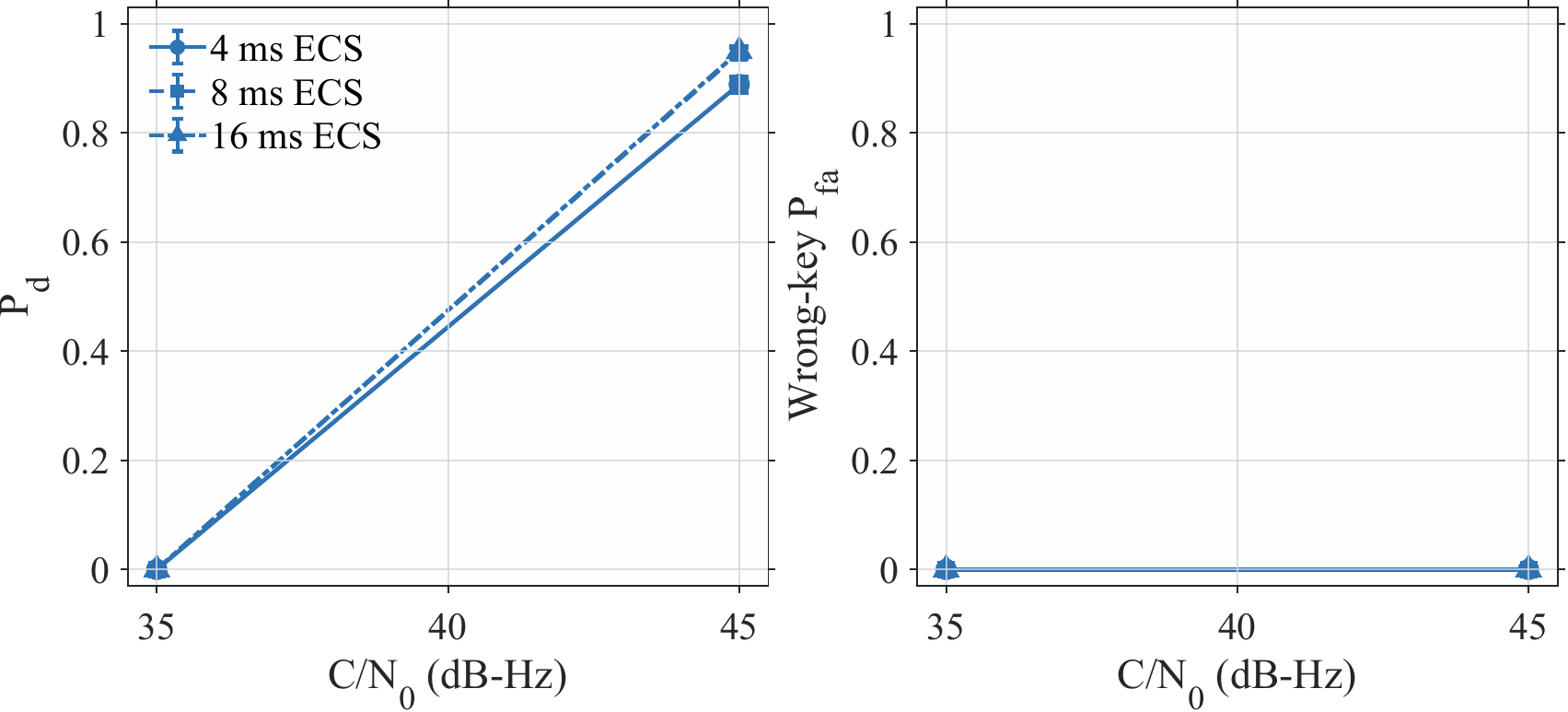}}
\caption{ECS detection probability and wrong-key false alarm probability under different C/N$_0$ values.}
\label{fig:cn0-detection-probability}
\end{figure}

\begin{figure*}[t]
\centerline{\includegraphics[width=0.95\textwidth]{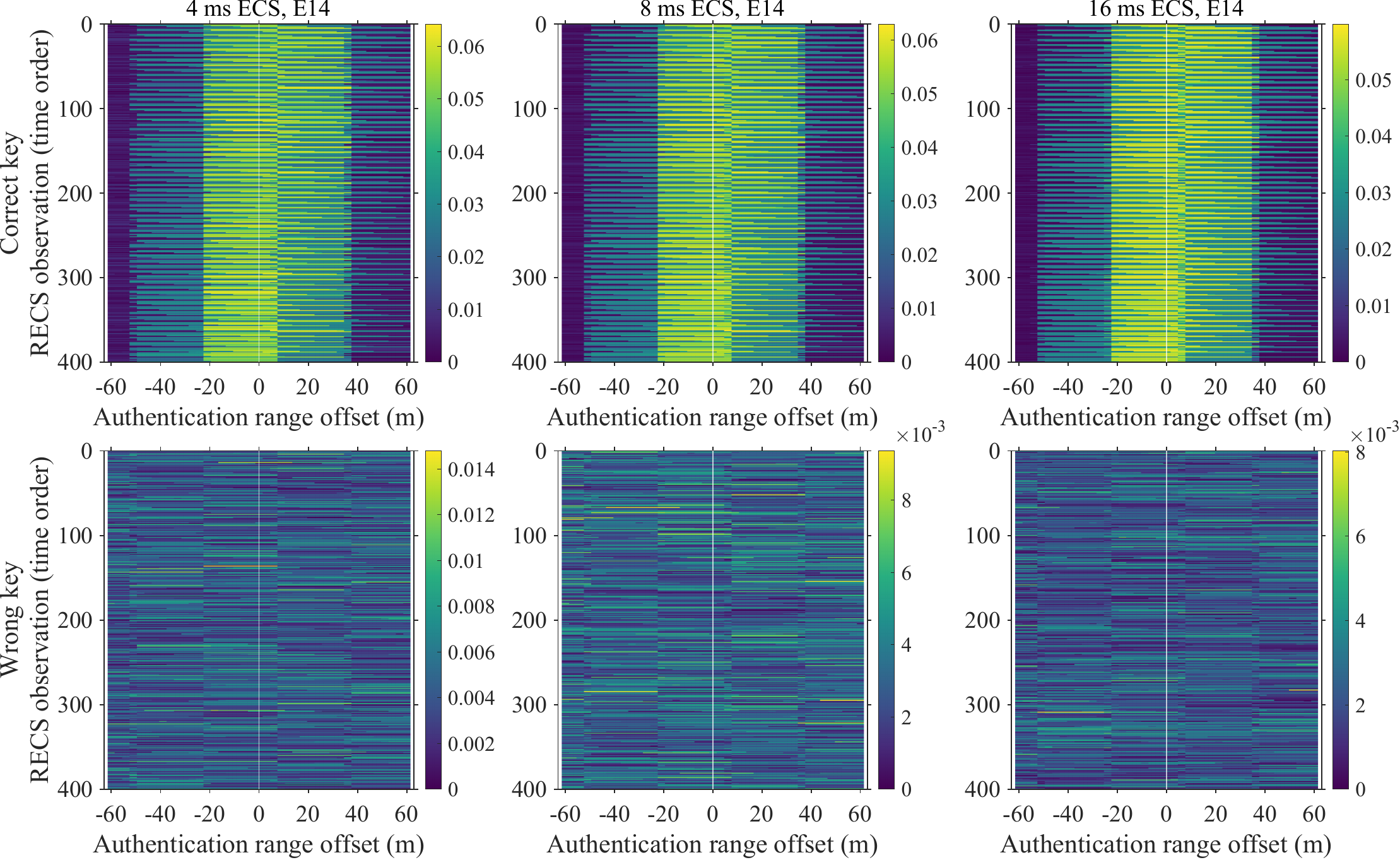}}
\caption{Per-record correlation waterfalls for correct-key and wrong-key trials at 45 dB-Hz.}
\label{fig:correlation-waterfall}
\end{figure*}

Window placement affects where the encrypted fragment is sampled within a RECS period, so it is important to verify that this scheduling choice does not bias the accepted correlation delay. 
Fig.~\ref{fig:window-position-range-bias} compares first, second, and last window placements for the 200-ms/16-ms configuration. 
The empirical distributions almost overlap, and the RMSE values remain close to 11.7--11.8 m. 
This suggests that, in the current synchronized simulation and receiver implementation, window placement can be selected for scheduling convenience without producing a clear range-bias penalty.

\subsection{Noise Robustness and Wrong-Key Control}

The previous experiments show that SAS works under clean conditions. 
We further evaluate whether the correlation-based decision remains meaningful when the E6-C snapshots are degraded by deterministic AWGN. 
The noise experiments fix the RECS period to 200 ms and the window position to the second window, then vary the ECS length and C/N$_0$.

Fig.~\ref{fig:cn0-detection-probability} shows a clear detection boundary. 
At 45 dB-Hz, the detection probability is 88.9\% for 4-ms ECS, 94.7\% for 8-ms ECS, and 95.0\% for 16-ms ECS. 
At 35 dB-Hz, none of the tested ECS lengths produces accepted records. 
Importantly, the wrong-key false alarm probability remains 0 in all tested noise configurations. 
This indicates that when signal is too weak, the receiver fails closed and rejects the record.

\begin{figure}[h]
\centerline{\includegraphics[width=\linewidth]{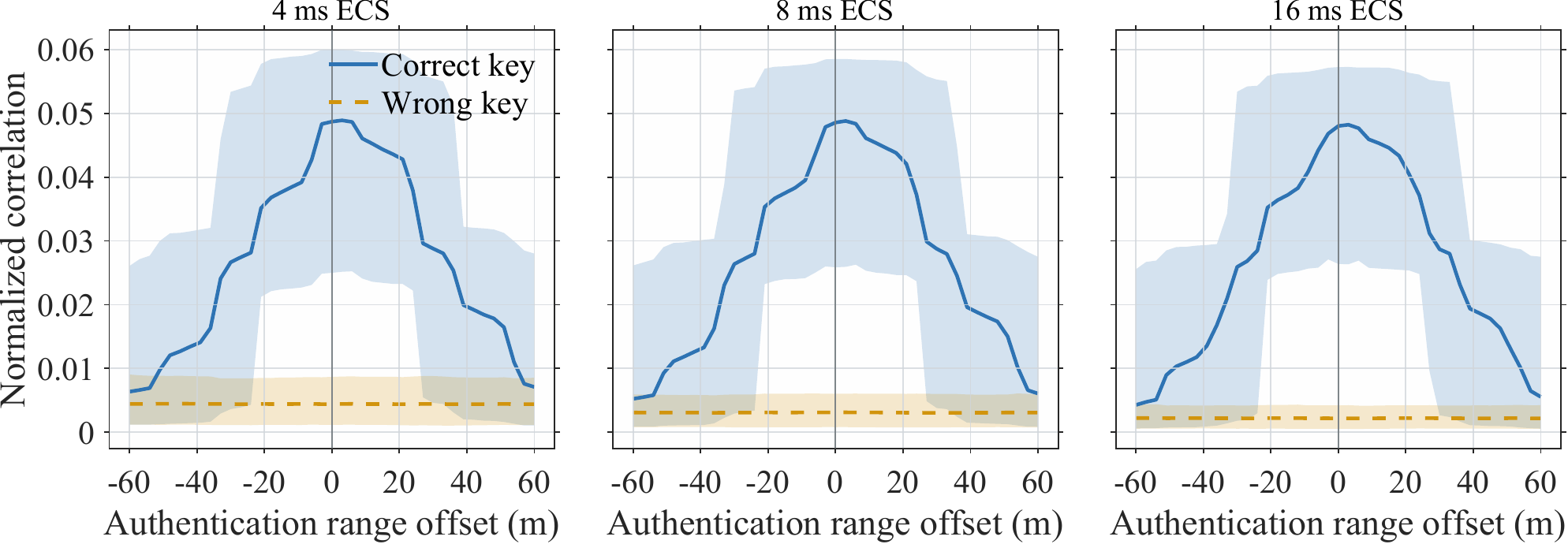}}
\caption{Mean correct-key and wrong-key correlation functions at 45 dB-Hz.}
\label{fig:correlation-curves}
\end{figure}

The correlation traces explain why the receiver can reject wrong keys. 
In Fig.~\ref{fig:correlation-curves}, the correct-key curves show a stable peak around the expected range offset, while the wrong-key control remains close to the noise floor. 
For example, at 45 dB-Hz the mean correct-key peak is about 0.055 for 4-ms ECS and about 0.054 for 8-ms and 16-ms ECS, whereas the corresponding wrong-key peaks remain around 0.007, 0.005, and 0.004. 
Fig.~\ref{fig:correlation-waterfall} further shows that this separation is not caused by a small number of outliers.
Correct-key observations repeatedly form an aligned high-correlation band, while the wrong key observations do not produce a persistent range-consistent structure. 
These results validate the receiver-side wrong-key comparison and show that RECS file metadata, key timing, and E6-C snapshot selection are being processed consistently.

\section{CONCLUSION}\label{sec:conclusion}
This paper presented GalSAS-SDR-SIM, an open-source SDR simulation platform for research on Galileo Signal Authentication Service. 
The platform integrates synchronized Galileo E1, E5b, and E6 signal generation, configurable OSNMA authentication, research-oriented E6-C code encryption.
By coupling E1 OSNMA and E6-C RECS through a shared TESLA key schedule, GalSAS-SDR-SIM supports end-to-end experiments in which navigation-message authentication and spreading-code authentication can be evaluated together.
Furthermore, the platform supports simulations at arbitrary locations and times, under both OSNMA and SAS configurations.
The evaluation shows that the platform can generate receiver-verifiable OSNMA data and use the verified E1 key timeline to drive SAS authentication. 
The configuration-sensitivity experiments further quantify how RECS period, ECS length, and window placement affect authentication rate, RECS file size, pending snapshot storage, processing time, and range-bias behavior.
Future work will extend SAS performance evaluation to more configurations and incorporate updated public SAS parameters.


\bibliographystyle{ieeetr}
\bibliography{ref}

\end{document}